# IRADCAL: A MONOLITHIC INORGANIC SCINTILLATOR AND THIN SCINTILLATORS TO MEASURE LOW ENERGY ELECTRON, PROTON AND HEAVY ION ALBEDO SPECTRUMS FROM LUNAR SURFACE

**A. B. Alpat[a]\*, A. Bozkurt[b], G. Bartolini[c], R. Bayram[d], A. I. Shah[e], L. Salvi[f], Y. Bakış[g], E. Hüseyinoğlu[h], T. Wusimanjiang[i], H. Raheem[j], D. Dolek[k], N. Ciccarella[l], S. Gigli[m]**

[a] Istituto Nazionale di Fisica Nucleare (INFN), Sezione di Perugia, Via A.Pascoli snc, 06123, Perugia, Italy, behcet.alpat@pg.infn.it
[b] IRADETS A.S., İstanbul Teknopark 4/A, Türkiye, arca.bozkurt@iradets.com
[c] BEAMIDE s.r.l., Via Campo di Marte 4/o, 06124, Perugia, Italy, giovanni.bartolini@beamide.com
[d] IRADETS A.S., İstanbul Teknopark 4/A, Türkiye, raziye.bayram@iradets.com
[e] BEAMIDE s.r.l., Via Campo di Marte 4/o, 06124, Perugia, Italy, ahmed.imam.shah@beamide.com
[f] Istituto Nazionale di Fisica Nucleare (INFN), Sezione di Perugia, Via A.Pascoli snc, 06123, Perugia, Italy, lucia.salvi@pg.infn.it
[g] IRADETS A.S., İstanbul Teknopark 4/A, Türkiye, yakup.bakis@iradets.com
[h] IRADETS A.S., İstanbul Teknopark 4/A, Türkiye, ersin.huseyinoglu@iradets.com
[i] Istituto Nazionale di Fisica Nucleare (INFN), Sezione di Perugia, Via A.Pascoli snc, 06123, Perugia, Italy, wusimanjiang.talifujiang@pg.infn.it
[j] BEAMIDE s.r.l., Via Campo di Marte 4/o, 06124, Perugia, Italy, haider.raheem@beamide.com
[k] IRADETS A.S., İstanbul Teknopark 4/A, Türkiye, deniz.dolek@iradets.com
[l] BEAMIDE s.r.l., Via Campo di Marte 4/o, 06124, Perugia, Italy, nora.ciccarella@beamide.com
[m] BEAMIDE s.r.l., Via Campo di Marte 4/o, 06124, Perugia, Italy, stefano.gigli@beamide.com
\* Corresponding Author

Presented at 75th International Astronautical Congress (IAC), Milan, Italy, 14-18 October 2024. IAC-24-A3,IP,193

**Abstract**

The Moon, lacking a magnetic field and a collisional atmosphere, is directly exposed to various space radiation types: Solar Wind (ions between 0.5 to 10 keV and lower energy electrons), Solar Energetic Particles (SEPs, ranging from 10 keV to several hundred MeV ions and electrons), Galactic Cosmic Rays (GCRs, with ions and electrons between 100 MeV and 10 GeV), and Anomalous Cosmic Rays (ACRs, ranging from 1 to 100 MeV particles). Monitoring SEPs and GCRs is critical to assess the lunar radiation environment in preparation for the return of humans to the Moon and to understand related radiation risks. When SEPs and GCRs reach the lunar surface, they can be absorbed, scattered, or remove atoms from the regolith, or create cosmogenic nuclides. Their interaction with the regolith also produces albedo energetic particles. As part of the Turkish Lunar Mission (TLM), a small acceptance particle detector is being developed to measure the albedo electron, proton, and heavy ion fluxes backscattered from the lunar surface. In low lunar orbit, the detector FoV (±10°) will look at the Moon surface. The IRADCAL detectors consist of several layers, from top to down: a Multi-Layer Insulator (MLI); a thin plastic scintillator (S1) seen by four Silicon Photomultipliers (SIPMs), placed to form a cross on light guide surrounding the scintillator (S1); a 26x26x70 mm$^3$ CsI(Tl) crystal scintillator (S2) and a thin crystal scintillator seen by four SIPMs (S3). The detector is designed to measure contained proton energy spectrum from 400 keV to 150 MeV, electron from 50 keV to 10 MeV and heavy ions (up to the CNO group) with energies up to 20 MeV/n. IRADCAL will provide two dE/dX values from two thin scintillators, $E_{tot}$ or dE/dX from thick scintillator, and dE/E ratios for particle identification. A multi-layer perceptron is being developed by using deep neural network algorithm to estimate the Depth of Interaction (DoI). S1 provides level 1 trigger and hit position, which, combined with DoI point, will help determine if a particle is escaping from the sides and to separate up/down-going particles. The thicker S3 will help identify whether the event is fully contained. It will fit in 2U CubeSat with 2.5 kg and 15 Watts of power required. It will also have capabilities such as remote control of discriminator threshold levels and of trigger logic, as well as automatic bias compensation, online calibration, and an onboard computer for data preprocessing.

**Nomenclature**
Performance tests [Pe], Qualification tests [Qu].

**Acronyms/Abbreviations**
Solar Energetic Particles (SEP), Galactic Cosmic Rays (GCRs), Anomalous Cosmic Rays (ACRs), Lunar Atmosphere and Dust Environment Explorer (LADEE), Lunar Reconnaissance Orbiter (LRO), SST (Solid State Telescope), Turkish Lunar Mission (TLM), Multi-Layer Insulator (MLI), Convolutional Neural Network (CNN), Particle IDentification (PID), Silicon PhotoMultiplier (SiPM), Field of View (FoV), Data Acquisition (DAQ), MicroProcessor Unit (MPU), Depth of Interaction (DoI), Fully Connected (FC), Analog to Digital Converter (ADC), Monte Carlo (MC).

## 1. Introduction

For about two-thirds of its orbit, the Moon is positioned outside Earth's magnetosphere, where its surface is exposed to various charged particles, such as low-energy solar wind plasma and high-energy SEPs (see Fig. 1). The latter are responsible for weathering and chemical changes on the lunar surface. During the remaining one-third of its orbit, the Moon is situated within Earth magnetosphere.

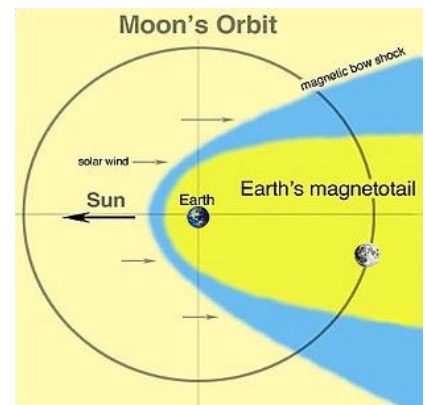

Fig. 1. Orbit of the Moon with respect to the magnetosphere of the Earth. The sizes of the Earth and of the Moon are not on scale. (Adapted from: Tim Stubbs / University of Maryland / GSFC) [1].

The Moon proximity offers an excellent setting for studying GCRs, solar wind, and SEPs. This environment closely resembles that of deep space, with the exception that the Moon acts as a barrier to GCRs and also interacts with them. For 5 to 6 days during each orbit, the Moon passes through the tail of Earth's magnetosphere, where it is shielded from the solar wind and instead exposed to the terrestrial magnetotail plasma environment. This offers a unique opportunity to study the dynamics of the magnetotail in situ and its relationship to solar and geomagnetic activity. The Moon is ideally positioned to observe atmospheric escape from Earth into space, specifically through the detection of energetic heavy ions rising from Earth's ionosphere and being carried into the distant magnetotail. Much of the current data on this phenomenon has been gathered by the THEMIS-ARTEMIS and SELENE spacecraft. The Moon surface-bounded exosphere, along with its production mechanisms, dynamics, and interactions with both the solar wind and terrestrial magnetotail plasma, as well as its eventual escape into space, are key areas of study. The LADEE and LRO have provided valuable info into the complexity of the lunar exosphere and the physical processes involved. Other fields of study are: energetic ion implantation into the lunar regolith, the production of albedo energetic particles from the interaction of SEPs and GCRs with the regolith, solar wind ion implantation or neutralization and reflection from the lunar regolith, formation of hydrogen-bearing molecules, potentially including water.

Since the Moon has not got an intrinsic magnetic field and a collisional atmosphere, it is directly exposed to Solar Wind (ions with energies ranging from approximately 0.5 to 10 keV and lower energy electrons), SEPs (ions and electrons with energies ranging from around 10 keV to several hundred MeV), GCRs (ions and electrons with energies ranging from about 100 MeV to around 10 GeV), ACRs (particles with energies ranging from about 1 to 100 MeV) (see Fig.2).

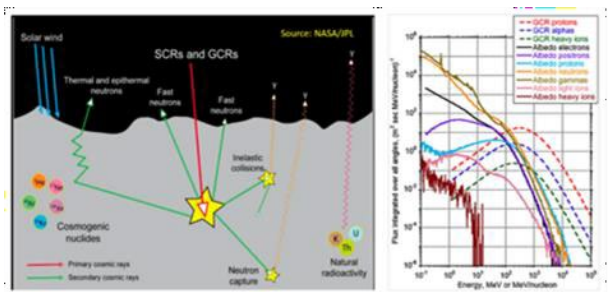

Fig. 2. Left: diagram illustrating the interaction of the solar wind and GCRs with the lunar regolith (Credit: NASA/JPL). Right: Energy spectra of pristine GCR species (dashed lines) and lunar albedo species (solid lines), modelled using the Geant4 simulation toolkit [1].

Monitoring SEPs and GCRs is crucial for assessing the Moon radiation environment, especially with plans for human return to the lunar surface and the associated radiation risks. When SEPs and GCRs impact the Moon surface they can be absorbed, scattered, they can remove a further atom from the lunar regolith, produce cosmogenic nuclides or produce albedo energetic particles because of their interaction with the lunar regolith.

The ARTEMIS mission, which began in April 2011 as an extension of the THEMIS mission (launched in 2008), includes two identical probes: P1 (THEMIS-B) and P2 (THEMIS-C).

Both probes carry the same set of instruments. These include FluxGate Magnetometer and ElectroStatic Analyzer for doing magnetic field and plasma measurements and SST (2008) to measure ions with energies ranging from 25 keV to 6 MeV and electrons having energies from 25 keV to 1 MeV.

From the results of THEMIS – ARTEMIS campaign, it outcomes that a key issue for next return of humans to the Moon is the shielding of astronauts and equipment on the lunar surface from energetic particle radiation. Even if the most dangerous damage to astronauts and electronics is caused by penetrating particles with energies greater than 1 MeV, the SEPs with energies above 25 keV can enter the magnetotail along open field lines that are more undesirable. Furthermore, the SEPs with energies from keV to MeV represent a risk during lunar surface exploration even if the Moon is inside the terrestrial magnetotail. Very high energetic particles (more than 10 MeV) are challenging to shield against, even with the protection offered by magnetosphere or the Moon itself. However, shielding against particles with lower energies but higher fluxes have not been examined yet [2, 3].

The Lunar Orbital Platform - Gateway is a crewed orbital platform that NASA, along with its international partners such as ESA, JAXA, and CSA, will assemble and it will operate near to the Moon. The platform will be equipped with comprehensive instrumentation for studying space plasma and energetic particles, spanning an energy range from approximately 5 eV for magnetospheric electrons to 5 GeV for GCRs [4].

As part of the TLM project (see Fig. 3), we are developing a compact particle detector payload to measure the fluxes of albedo electrons, protons, and heavy ions that are backscattered from the lunar surface.

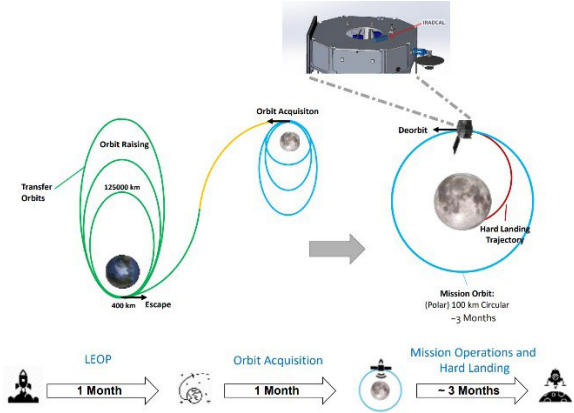

Fig. 3. Overview of the TLM mission and IRADCAL put on the satellite [5, 6].

In low lunar orbit, our payload will have a FoW (+/- 10 degrees) directed towards the Moon surface. A prolonged stay on the Moon would expose the human body to significantly high levels of radiation because in space the radiation is usually much more intense than on Earth. Before humans return to the Moon, IRADCAL [7] will be among the few radiation detectors collecting data during the journey to the Moon and after landing on its surface. IRADCAL will be part of the TLM mission. The data gathered by IRADCAL will offer important information regarding space weather conditions both in orbit and on the Moon.

## 2. IRADCAL detector and performance

The parameters of the design are: energy spectrum for contained protons (from 400 keV to 150 MeV), electrons (from 50 keV to 10 MeV) and heavy ions up to CNO group (50 MeV/n); two dE/dx values given by the two thin scintillators; $E_{dep}$ or dE/dx from the thick scintillator and dE/E that is really important to recognize particles [8]. The main parameters of IRADCAL are reported in Table 1.

Table 1. The most important parameters of IRADCAL

| Features | Specification |
|---|---|
| Size | Fits in 2U CubeSat (100x100x200 mm$^3$). Standard 2U Cubesat modified to be mounted on satellite |
| FoV | ≃ ±10 degrees |
| Mass | ≃ 2.5 kg |
| Power | ≃ 15.0 Watts, 25-34 V (nom. 28 V) |
| Communication | RS-422 Interface with Satellite |
| Downlink/Uplink limit | 10 MBytes/24h, 1000 bit/s per operation. Data compression onboard |
| Temperature Range | -20 to + 40°C in operation with ability to withstand -20°C to + 60°C when non-operational |
| Slow Control | Temperature of electronics, leakage current of SiPMs, remote control of discriminator threshold levels, remote control of trigger logic, automatic bias compensation, online calibration capability |

*2.1 Detector*

The IRADCAL detectors consist of the following components, from top to bottom: a MLI, followed by a thin plastic scintillator (S1) read by four SIPMs arranged to form a cross on light guide surrounding the scintillator, a 12x12x70 mm$^3$ CsI(Tl) crystal scintillator (S2), and, at the bottom, a thin crystal scintillator (S3) read by two SIPMs (see Fig. 4).

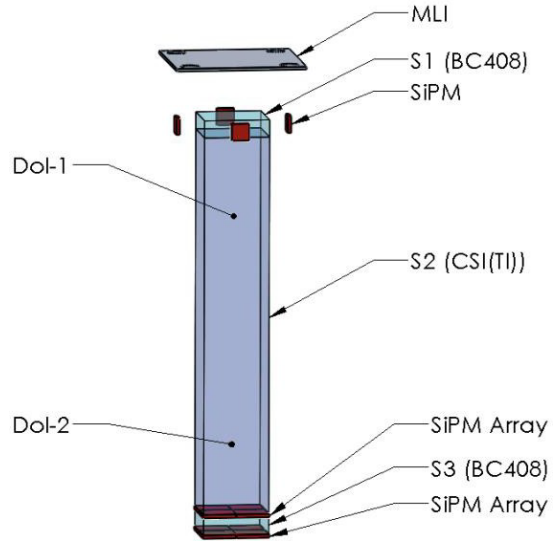

Fig. 4. Scheme of IRADCAL detectors.

In more details the constituents are:
- thin plastic Scint-1 (S1), it gives the Level 1 Trigger and $dE_1$ of charged particles. Counts the total number of photons on 4 SiPMs;
- thick crystal scintillator (CsI(Tl) (S2)), it gives E (contained = stopping), dE (not contained = not stopping) of charged and neutral particles. Counts the total number of photons on 4 SiPMs;
- thin crystal Scint-2 (S3), it gives $dE_2$ of charged and neutral particles. Counts the total number of photons on 2 SiPMs.

The information is combined on an event-by-event basis and the signals from the SiPMs are 10 in total. The values of $dE_1$, E, $dE_2$, the ratios $dE_1/E$ and $dE_2/E$, together with the CNN are employed to define: on $dE_1$ and $dE_2$ weighted impact positions contained events with their deposited energies, PID (e, p, HI), $DoI_1$ and $DoI_2$ x,y,z coordinates, veto on side impact events, up and down separation.

*2.2 Mechanic Interface*

The IRADCAL detector and associated electronics are designed to integrate into standard 2U CubeSat [9] structures (see Fig. 5).

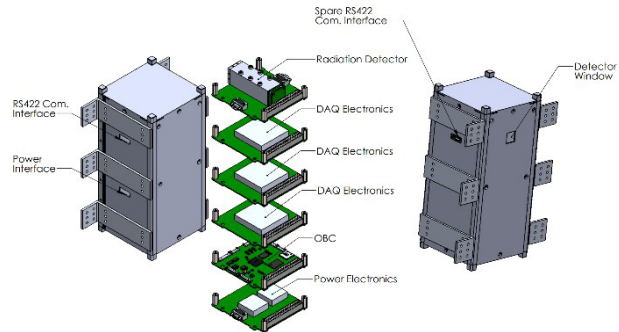

Fig. 5. Exploded view of IRADCAL.

For use as a scientific payload in TLM, the CubeSat structure has been modified with a mounting interface.

*2.3 Electronics*

The DAQ electronics for the IRADCAL system will consist of amplifier, peak detector, comparator, and ADC circuits, all of which will be controlled by an MPU. The primary objective of the DAQ system is to capture both the timing and energy of optical photons generated in the scintillator. Given that the current produced by these optical photons is on the order of nanoamperes and lasts for only a few hundred nanoseconds, custom DAQ electronics are required to meet these performance demands.

The optical signals from the SiPM will be amplified at a fixed gain using an amplifier circuit. The amplified signal will then be input to the peak detector and comparator circuits. The peak detector will store the maximum value of the amplified signal until it is ready to be processed by the ADC. Meanwhile, the comparator will evaluate the amplified signal against a predefined threshold; if the signal exceeds this threshold, it signifies an optical event at the SiPMs. The comparator will then generate a trigger signal for the MPU.

This process will occur asynchronously across all SiPMs and DAQ channels. The MPU will apply coincidence logic to the trigger signal it receives to suppress dark counts. If an event passes the coincidence logic, then MPU will command the ADCs to measure the energy on all channels. All timestamps and ADC readings will be processed and compressed by the MPU. Results will be saved on the SD Card. Upon a request from the satellite, the MPU will send the data over an RS422 interface for downlinking.

IRADCAL is powered by the satellite's nominal 28 V power supply, which is regulated down to 3.3 V to support internal electronics. This regulation is achieved using a space-qualified EMI filter and DC-DC conversion module. The regulated power is then distributed to all electronic circuits via the PC104 interface. Additionally, a low-noise DC-DC regulator with proven space heritage is implemented to power the SiPMs.

For slow control, communication interfaces, and DAQ management, the IRADCAL system utilizes the space-qualified NSTOBC from Nara Space. The MPU responsible for DAQ operations is also located in the OBC. Communication with the satellite, including the transmission of downlink data, is facilitated via RS422 modules on the OBC [10].

**3. New method to select only "Contained Events"**

In order to identify particles thanks to their energy released in the three scintillators, it is necessary to select only down-going events that hit S1 and stop in S2. To subtract the background, that is given by events going from down to up and by events hitting S1 and S2 but escaping from the S2 side (so the event is not contained), a DoI prediction model has been developed (see Fig. 6).

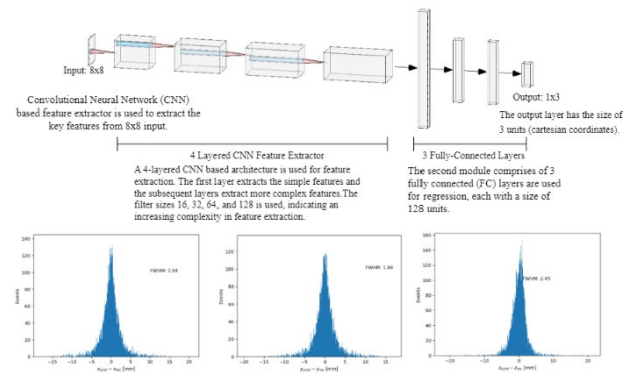

Fig. 6. In the upper image the DoI prediction CNN model is shown. In the lower part, from the left the difference in millimeters between the real and model reconstructed DoI coordinates.

It calculates the coordinates of the first and last point belonging to a single track. The architecture of the DoI Prediction model consists of two principal modules: a CNN based feature extractor and FC layers. The first module is employed to extract the key features from 8x8 input. Four convolutional layers with filter sizes 16, 32, 64 and 128 are employed. This shows an increasing complexity in feature extraction. The second module, instead, is composed by three FC layers that are employed for regression. Each one has a size of 128 units and output layer with a size of 3 units (cartesian coordinates for the start or the end of the interaction).

To implement the CNN Model, it is necessary to have extensive data training first with laboratory sources, then at particle accelerators (see Fig. 7).

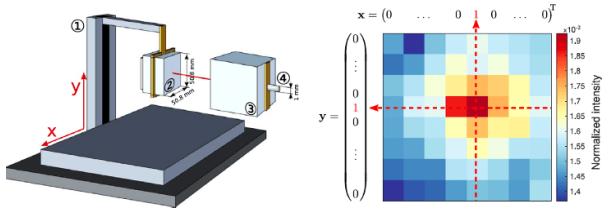

Fig. 7. Motorized stage used to validate MC simulations with a laboratory source scanning system.

So, the following approach was used:
- extensive dataset creation (80% training and 20% for modelling; impact angle, energy, particle type and DoIs x,y,z) thanks to MC simulation in order to improve CNN algorithm;
- IRADCAL data collection on motorized test bench with distinct impact angles and energies for beta particles from laboratory sources, comparison with data from MC simulations and validation of the CNN algorithm;
- collection of training and real data by bringing IRADCAL to electron and proton accelerators.

## 4. Simulations results and Discussion

From the simulations, the protons and electrons generated and deposited energies can be plotted for the case S3 = 0, no DoI cut. Also, the protons escaping from S2, beginning at a kinetic energy of nearly 160 MeV, can be charted (see Fig. 8).

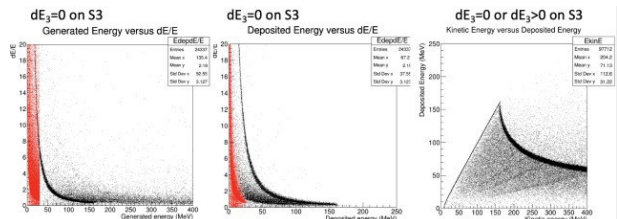

Fig. 8 From left, electrons and protons generated and deposited energies; protons escaping from S2.

The beam tests to study the performance and the radiation qualification that we will carry out are the following ones:
- EMI/EMC Tests (TBD);
- Mechanical Test, in particular, regarding vibration, acceleration and shock (Tubitak Uzay);
- TVAC Tests (Tubitak Uzay).

More details regarding the tests of the performance and qualification against the space radiation are reported in Table 2.

Table 2. Tests for the performance and qualification against space radiation

| # | Beam | ECSS | Description | Facility |
|---|------|------|-------------|----------|
| 1[Pe] | e- | N/A | 100 keV - 10 MeV | TBD |
| 2[Pe] | p | 22900, 25100 | 8 – 30 MeV | TENMAK |
| 3[Qu] | $^{60}$Co | 22900, 25100 | 1.17- 1.33 MeV | TENMAK |
| 4[Pe] | $^{12}$C | 22900, 25100 | 10 MeV/n | UCL-Belgium |

*4.1 Three-point linear fit strategy*

A simple linear fit is used to combine three points into a track. The three points are the position in the first thin scintillator (S1), and the two DoI points in the main scintillator (calorimeter) for first and last scintillation positions. These results are produced using 300 thousand simulated events for proton particles from 200 keV to 200 MeV, arriving from above at different angles.

The three points fit is done separately in the xy plane and zy plane. The Chi2 value is lower than 1 in most of the events, meaning the assumption that tracks follow a linear path can be considered correct (see Fig. 9).

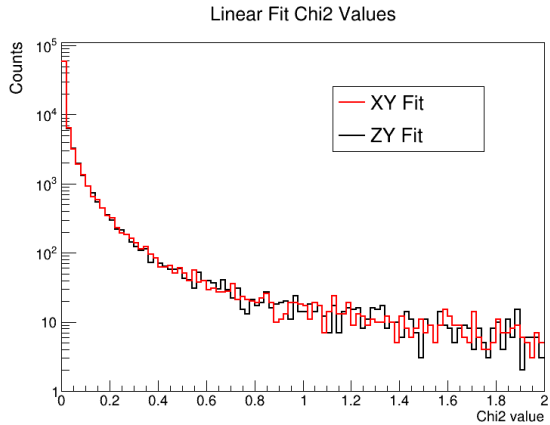

Fig. 9 Plot of the Chi2 values for the XY plane linear fit (red), and for the ZY plane (black).

Using the fitted track, the exit position of the incoming particle can be evaluated. A simple event rejection strategy (Base Strategy) is applied by rejecting events which have the fitted track pointing to the side of the Calorimeter (Fit-Exited), and only accepting events with the fitted track pointing to the bottom (Fit-Contained), that do not have signal in the second thin scintillator. However, this was found to have a very low event acceptance, with less than 10% of truth-contained events that are accepted.

To improve the acceptance of the event selection strategy, two simple cuts based on the last DoI position were added (Improved Strategy). If the last interaction position is more than 2 mm far from the lateral sides of the Calorimeter, the event is considered contained. Also, if the last DoI position is less than 4 mm far from the bottom, but the event doesn't have signal in the second thin scintillator, the event is again considered contained. After adding these cuts to the event rejection strategy, the contained event acceptance improved from less than 10% to over 50%, even if also the false positive events increased from less than 1% to more than 6% of the truth-exited events (see Fig. 10).

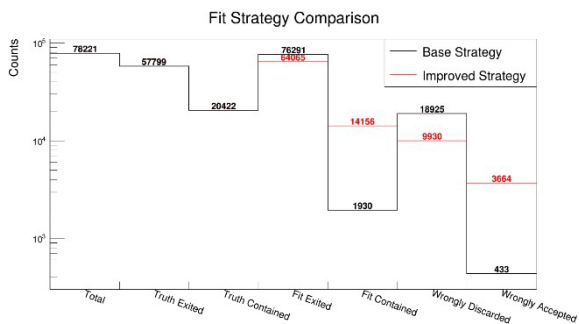

Fig. 10 Comparison between Base Strategy and Improved Strategy.

The values used in the two cuts of the improved strategy were just taken as a demonstration and are not yet optimized. Furthermore, more complex strategies involving more complex algorithms can be studied to increase the performance of the event selection.

*4.2 Particle Identification Improvement using DoI event selection*

The presented DoI-based event selection strategy can be helpful to improve the particle identification efficiency. A simulation including protons from 200 keV to 200 MeV, electrons from 10 keV to 10 MeV and $^{12}C$ ions with 50 MeV/n energy, with particles arriving from above at different angles, was used to study the effect of the event selection strategy.

As shown in the Calorimeter energy deposit plots, after adding the DoI-based selection (see Fig. 11) the overlap between electrons and protons is significantly reduced at low energy values. This increase in the separation of the two distributions can improve significantly the particle identification efficiency of the final IRADCAL system.

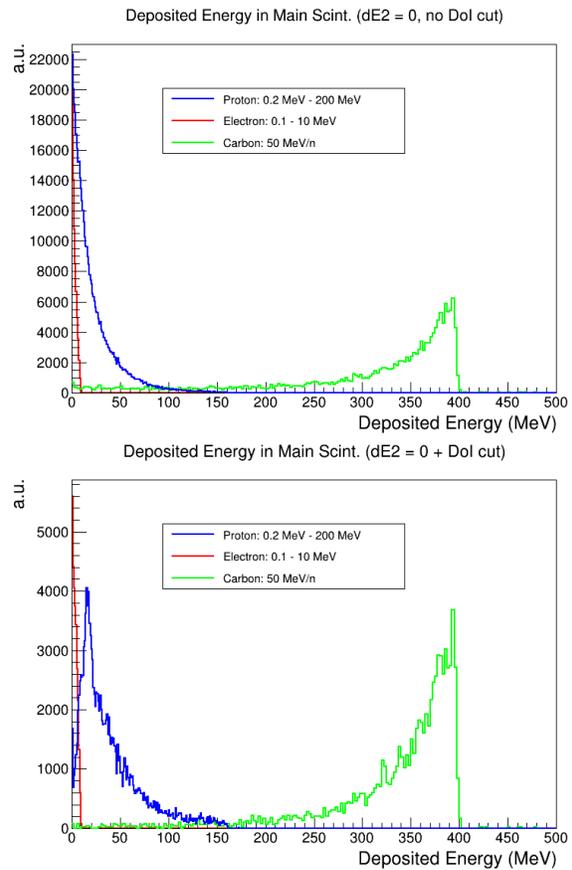

Fig. 11 In the first picture, calorimeter energy deposit without the DoI cut; in the second picture, calorimeter energy deposit with DoI cut.

*4.3 Preliminary Background Studies*

A preliminary study on the main background components, which are gamma and neutron particles, has been performed. Gamma particles between 150 keV and 15 MeV are simulated and compared with the electron simulation, using the same number events (200 thousand). The same event selection requirements of energy deposit > 0 on both the first thin scintillator and the Calorimeter, as well as the DoI-based cut is applied to both sets, with an electron-gamma efficiency ratio of almost 50. The same procedure was applied to 300 thousand simulated events of neutron particle from 0.2 to 200 MeV and compared with the same number of simulated proton events. After applying the full event selection requirements to both sets, a proton-neutron efficiency ratio of around 40 is obtained (see Fig. 12).

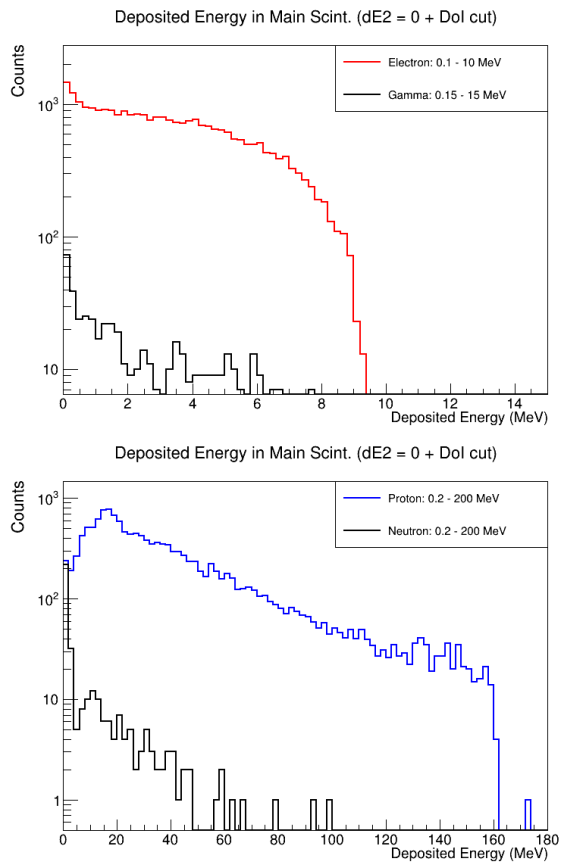

Fig. 12 In the first picture, gammas and electrons comparison; in the second picture, neutrons and protons comparison.

These results are very promising as they show that the event selection efficiency for charged particles is much higher than the efficiency on background events, which will benefit from the final signal to noise ratio. A more appropriate background study needs to be performed by considering the expected flux of each particle type at the relevant energies, to properly estimate the signal to noise ratio of the event selection [11].

## 5. Conclusions

IRADCAL will be the first charged particle detector demonstrator performing calorimetry designed and built in Turkey for space applications. It is currently under construction by IRADETS A.S., in collaboration with BEAMIDE srl (a spin-off company of INFN), and it is supported by Tubitak-Uzay. The project is expected to be completed by July 2025. IRADCAL is a compact, lightweight, and low-power detection system capable of free-flying as a CubeSat (also in constellation) or it can be mounted on larger satellites.

It will gather valuable scientific data on low energy electrons, protons and heavy ion fluxes and energies during its journey to the Moon and while looking at to its surface. This information is important in view of near future manned missions to the Moon. In addition to its scientific role, IRADCAL will serve as a radiation hazard alarm system, protecting satellites from stochastic radiation hazards by enabling real-time on-board mitigations and sending alerts to Earth. Detailed simulations using the MRADSIM tool [12] are currently in progress.

**Acknowledgements**

The authors would like to express their gratitude to TÜBİTAK Uzay and the Turkish Space Agency for accepting IRADCAL as a scientific payload in the Turkish Lunar Mission and for providing the necessary facilities and resources for its development. Special thanks to Burak Yağlıoğlu for his invaluable guidance and insightful discussions throughout this project. Additionally, we acknowledge Burak Karagözoğlu and the rest of the Tubitak Uzay team for their technical assistance and support. Finally, we extend our appreciation to NaraSpace for their assistance with the mechanical and control interfaces.